\documentclass[aps,showkeys,showpacs,superscriptaddress,a4paper,10pt]{revtex4-2}

\usepackage{amssymb}
\usepackage{amsmath}
\usepackage[dvips]{graphicx}
\usepackage[T2A]{fontenc}
\usepackage{graphicx}
\usepackage[breaklinks=true,colorlinks=true,linkcolor=blue,urlcolor=blue,citecolor=blue]{hyperref}
\usepackage{comment}

\begin{document}

\title[Cahn-Hilliard/Allen-Cahn Equation with memory effects]{Convective Viscous Cahn-Hilliard/Allen-Cahn Equation with memory effects}

\author{P.~O.~Mchedlov-Petrosyan}
\author{L.~N.~Davydov}
\email[Corresponding author: ]{ldavydov@kipt.kharkov.ua}

\affiliation{National Science Center "Kharkiv Institute of Physics and Technology",   \\
1, Akademichna St., Kharkiv, 61108, Ukraine}

\begin{abstract}
The combination of the well-known Cahn-Hilliard and Allen-Cahn equations is used to describe surface processes, such as simultaneous adsorption/desorption and surface diffusion. In the present paper we have considered the convective-viscous Cahn-Hilliard/Allen-Cahn equation complemented by memory effects. Exact solutions are obtained and the combined action of the applied field, dissipation and memory are discussed.
\end{abstract}


\maketitle

\renewcommand{\theequation}{\arabic{section}.\arabic{equation}}
\section{Introduction}\label{s1}

The present work is devoted to the study of the equation for the order parameter $\bar{w}$:
\begin{equation} \label{1.1} \frac{\partial \bar{w}}{\partial t'} -2\bar{\alpha }\bar{w}\frac{\partial \bar{w}}{\partial x'} +\tau '\frac{\partial }{\partial t'} \left(\frac{\partial \bar{w}}{\partial t'} +M_{2} \bar{\mu }\right)=\frac{\partial }{\partial x'} \left[M_{1} \frac{\partial }{\partial x'} \left(\bar{\mu }+\bar{\eta }\frac{\partial \bar{w}}{\partial t'} \right)\right]-M_{2} \bar{\mu }. \end{equation}
This is a one-dimensional version of the Convective-Viscous Cahn-Hilliard/Allen-Cahn equation (CVCH/AC), modified by memory effects in diffusion. The combination of the well known Cahn-Hilliard (CH) and Allen-Cahn (AC) models was introduced in \cite{1} to describe surface processes, such as simultaneous adsorption/desorption and surface diffusion.
Here $\bar{\mu }$ is the chemical potential, $\bar{\alpha }$ is proportional to the applied field, $\bar{\eta }$ is the viscosity, and $\tau '$ is the characteristic memory time. $M_{1} $ is the mobility and $M_{2} $ kinetic coefficient. The chemical potential $\bar{\mu }$ is a variational derivative of the double-well thermodynamic potential $F\left[\bar{w}\right]$, which contains $\left(\frac{\partial \bar{w}}{\partial x'} \right)^{2} $ term and a fourth-power polynomial:
\begin{equation} \label{1.2} \bar{\mu }=-\bar{\varepsilon }\frac{\partial ^{2} \bar{w}}{\partial x'^{2} } +\bar{\rho }\left(\bar{w}-\bar{a}_{1} \right)\left(\bar{w}-\bar{a}_{2} \right)\left(\bar{w}-\bar{a}_{3} \right). \end{equation}
The roots $\bar{a}_{1} <\bar{a}_{2} <\bar{a}_{3} $ of the polynomial correspond to the positions of two minima and a single intermediate maximum of the fourth-power polynomial in the thermodynamic potential; $\bar{\varepsilon }$ is usually presumed to be of the order of the capillary length squared.

As it was pointed out by Thiele \cite{2}, the combination of CH and AC equations could be considered as a rather general description of the time evolution of a dissipative system, combining conserved and non-conserved dynamics of the order parameter. Without memory effects (which will be discussed below) the time evolution is governed by
\begin{equation} \label{1.3)} \frac{\partial \bar{w}}{\partial t'} =\nabla \left[M_{1} \left(\bar{w}\right)\frac{\delta F}{\delta \bar{w}} \right]-M_{2} \left(\bar{w}\right)\frac{\delta F}{\delta \bar{w}} . \end{equation}
Here $F\left[\bar{w}\right]$ is the thermodynamic potential functional; in the present work $M_{1} ,\, M_{2} $ are presumed to be constants.

Now, to understand the meaning of different terms in our modification \eqref{1.1}-\eqref{1.2} of this equation, we need to give some insight into the history of CH and AC equations.

Starting with the CH equation, we give only brief explanation of some basic assumptions and description of the model modifications; we also refer to the original papers \cite{3,4} and excellent reviews \cite{5,6}. The basic underlying idea of the Cahn-Hilliard model \cite{3,4} is that for an inhomogeneous system, e.g. a system undergoing a phase transition, the thermodynamic potential (e.g. free energy) should depend not only on the order parameter, but on its gradient as well. For the thermodynamic potential, containing the square of the gradient, the local chemical potential $\bar{\mu }$, defined as a variational derivative of the thermodynamic potential functional, contains the Laplacian. In our one-dimensional version it is the second derivative, see \eqref{1.2}. If the diffusional flux is proportional to the gradient of the chemical potential (Fickean diffusion), the diffusion equation for the order parameter is of the fourth order. The classic Cahn-Hilliard equation was introduced as early as in 1958 \cite{3,4}; the stationary solutions were considered, the linearized version was treated and corresponding instability of the homogeneous state identified. However, intensive study of the fully nonlinear form of this equation started essentially later \cite{7}. Now an impressive amount of work is done on the nonlinear Cahn-Hilliard equation, as well as on its numerous modifications, see \cite{5,6}. An important modification was done by Novick-Cohen \cite{8}. Taking into account the dissipation effects which are neglected in the derivation of the classic Cahn-Hilliard equation, she introduced the \textit{viscous} Cahn-Hilliard (VCH) equation, with the term $\bar{\eta }\frac{\partial \bar{w}}{\partial t'} $ added to the chemical potential $\bar{\mu }$. To account for the external field several authors considered the nonlinear \textit{convective} Cahn-Hilliard equation (CCH) \cite{9,10,11}. To study the joint effects of nonlinear convection and viscosity, Witelski \cite{12}[12] introduced the \textit{convective-viscous}-Cahn--Hilliard equation (CVCHE) with a general symmetric double-well potential $\Phi \left(\bar{w}\right)$:
\begin{equation} \label{1.4} \frac{\partial \bar{w}}{\partial t'} -\bar{\alpha }\bar{w}\frac{\partial w}{\partial x'} =\frac{\partial }{\partial x'} \left[M_{1} \frac{\partial }{\partial x'} \left(\bar{\mu }+\bar{\eta }\frac{\partial \bar{w}}{\partial t'} \right)\right] ,\end{equation}
\begin{equation} \label{1.5)} \bar{\mu }=-\bar{\varepsilon }\frac{\partial ^{2} \bar{w}}{\partial x'^{2} } +\frac{d\Phi \left(\bar{w}\right)}{d\bar{w}} . \end{equation}

It is worth noting that all results, including the stability of solutions, were obtained without specifying a particular functional form of the potential. Thus, they are valid both for the polynomial and logarithmic \cite{5,6} potential. Also, with an additional constraint imposed on the nonlinearity and viscosity, the approximate traveling-wave solutions were obtained. In \cite{13}, for equation \eqref{1.4} with polynomial potential, see \eqref{1.2}, several exact single- and two-wave solutions were obtained. Generally, the CH equation with convective and viscous terms appeared to be a useful tool in the study of further generalizations, such as external nonlinear sink/source terms and memory effects \cite{14,15}.

Evidently, the Cahn-Hilliard equation describes the evolution of the locally conserved order parameter. For the evolution of the non-conserved order parameter Allen and Cahn \cite{16} proposed the equation
\begin{equation} \label{1.6)} \frac{\partial \bar{w}}{\partial t'} =-M_{2} \left[-\bar{\varepsilon }\Delta \bar{w}+f\left(\bar{w}\right)\right]. \end{equation}
That is, the change of the order parameter is proportional to the deviation of the local chemical potential \eqref{1.2} of an inhomogeneous system from its equilibrium value.

The catalysis and deposition from liquid and gaseous phase involve, besides the overall mass transport, a manifold of processes at the surface itself, e.g. attachment/detachment of atoms, surface diffusion, chemical reactions, etc. Traditionally these surface processes were modeled by continuum-type reaction-diffusion models, see, e.g., \cite{17}. Such an approach usually neglects detailed interaction between the atoms. A different model was developed in \cite{1} and studied in the series of papers \cite{18,19,20,21,22,23}. On the microscopic level the basis of this approach is the modeling of surface processes by employing dynamic Ising type systems. Then the coarse-grained, mesoscopic models of Ising systems were derived \cite{24,25,26,27}. As the next step, to illustrate the effects of multiple mechanisms, the simplification of the mesoscopic equation (which retains its fundamental structure) for the order parameter $\bar{w}$ was introduced \cite{1,18}:
\begin{equation} \label{1.7} \frac{\partial \bar{w}}{\partial t'} =\frac{\partial }{\partial x'} \left(M_{1} \frac{\partial \bar{\mu }}{\partial x'} \right)-M_{2} \bar{\mu },   \end{equation}
which is a composition of the CH and AC equations. Also a more general nature of the latter equation was then noticed \cite{2}. It was found that the thin-film-type equations also fit the form \eqref{1.7}, if due to some mechanism the mass is not conserved \cite{2,28,29}.

If the conserved dynamics is described by the {\textit{convective-viscous} Cahn-Hilliard equation, the CH/AC equation, instead of \eqref{1.7}, is
\begin{equation} \label{1.8} \frac{\partial \bar{w}}{\partial t'} -2\bar{\alpha }\bar{w}\frac{\partial \bar{w}}{\partial x'} =-\frac{\partial J}{\partial x'} -M_{2} \bar{\mu }. \end{equation}
Here $J$ is the diffusional flux
\begin{equation} \label{1.9} J=-M_{1} \frac{\partial }{\partial x'} \left(\bar{\mu }+\bar{\eta }\frac{\partial \bar{w}}{\partial t'} \right). \end{equation}

The latter equation is the `generalized Fick's law' (modified by the presence of the viscous term). The Ficks law was often criticized for the infinite velocity of the spread of the diffusing substance. The most popular alternative is the Maxwell-Cattaneo approach, see \cite{30} for a comprehensive discussion; here we give only the one-dimensional formulae, in accordance with the spirit of the present paper. In the Maxwell-Cattaneo approach the mass-conservation equation \eqref{1.8} is not changed. However, instead of \eqref{1.9}, the following relation is proposed:
\begin{equation} \label{1.10} \tau '\frac{\partial J}{\partial t'} +J=-M_{1} \frac{\partial }{\partial x'} \left(\bar{\mu }+\bar{\eta }\frac{\partial \bar{w}}{\partial t'} \right). \end{equation}
Direct integration of the latter expression for the flux yields
\begin{equation} \label{1.11)} J=-\int _{0}^{t'}\left[\frac{M_{1} }{\tau '} \frac{\partial }{\partial x'} \left(\bar{\mu }+\bar{\eta }\frac{\partial \bar{w}}{\partial t'} \right)\right] \exp \left(\frac{t''-t'}{\tau '} \right)dt''. \end{equation}

So, this approach is also called `diffusion with memory effects'. Correspondingly, $\tau '$ is considered as the characteristic time of memory. On the other hand, eliminating $J$ from \eqref{1.8} and \eqref{1.10} yields \eqref{1.1}, i.e., the `hyperbolic modification' of the convective-viscous CH/AC equation. The hyperbolic modification of the classical Cahn-Hilliard equation was proposed in \cite{31} to model rapid spinodal decomposition in a binary alloy. However, from purely mathematical point of view -- as a singular perturbation of the classic Cahn-Hilliard equation -- it was considered even earlier \cite{32}. These papers where followed by many others, both of physical and mathematical nature \cite{33,34,35,36,37}. For the hyperbolic convective-viscous Cahn-Hilliard equation the exact traveling-wave solution was given in \cite{13}.

As discussed above, the CH and AC `parts' represent the different mechanisms involved in the process. Competition of these mechanisms may result in a quite complicated behavior, as found in the above cited papers \cite{2,18,19,20,21,22,23,29}. So it may be interesting to ask, when this mechanisms are `cooperative', i.e., result, e.g., in a single constant-velocity traveling wave of the phase transformation. Some preliminary results on traveling wave solution for the CH/AC equation without memory effects were given in \cite{38}. In the next Section \ref{s2} we find the exact traveling-wave solution for \eqref{1.1}. In Section \ref{s3} we consider the parametric dependence of this solution and discuss our results.

\setcounter{equation}{0}
\section{Traveling wave solution}\label{s2}

Introducing the non-dimensional order parameter $w=\frac{\bar{w}}{w_{0} } $, $w_{0} =\frac{1}{\sqrt{\rho } } $, non-dimensional coordinate $x=\frac{x'}{X} $ and non-dimensional time $t=\frac{t'}{T} $, we rewrite equations \eqref{1.1}-\eqref{1.2} in the non-dimensional form, where it is convenient to take $X=\sqrt{\bar{\varepsilon }} $; $T=\frac{\bar{\varepsilon }}{M_{1} } $,
\begin{equation} \label{2.1} \frac{\partial w}{\partial t} -2\alpha w\frac{\partial w}{\partial x} +\tau \frac{\partial }{\partial t} \left(\frac{\partial w}{\partial t} +h\mu \right)=\frac{\partial ^{2} }{\partial x^{2} } \left(\mu +\eta \frac{\partial w}{\partial t} \right)-h\mu  .  \end{equation}
\begin{equation} \label{2.2} \mu =-\frac{\partial ^{2} w}{\partial x^{2} } +\left(w-a_{1} \right)\left(w-a_{2} \right)\left(w-a_{3} \right) .  \end{equation}
We also introduced the notations
\begin{equation} \label{2.3)} \mu =\frac{\bar{\mu }}{w_{0} } ;\, \, \alpha =\bar{\alpha }\frac{w_{0} T}{X} ;\, \, T=\frac{X^{2} }{M_{1} } ;\, \, \eta =\frac{\bar{\eta }}{T} ;\, \, h=M_{2} T;\, \, \tau =\frac{\tau '}{T} ;\, \, w_{0} =\frac{1}{\sqrt{\bar{\rho }} } . \end{equation}

Changing variables $x,\, t\, \Rightarrow z,\, t$, where $z=x-vt$,  we get
\begin{eqnarray} \label{2.4}  &&{\frac{\partial w}{\partial t} -v\frac{\partial w}{\partial z} -2\alpha w\frac{\partial w}{\partial z} +\tau \frac{\partial }{\partial t} \left(\frac{\partial w}{\partial t} -v\frac{\partial w}{\partial z} +h\mu \right)} \nonumber \\ &&{-\tau v\frac{\partial }{\partial z} \left(\frac{\partial w}{\partial t} -v\frac{\partial w}{\partial z} +h\mu \right)=\frac{\partial ^{2} }{\partial z^{2} } \left[\mu +\eta \left(\frac{\partial w}{\partial t} -v\frac{\partial w}{\partial z} \right)\right]-h\mu }  .\end{eqnarray}
Looking for the traveling wave solutions we set $w=u\left(z\right)$; then \eqref{2.4} and \eqref{2.2} take the forms
\begin{equation} \label{2.5} -\frac{d}{dz} \left[vu+\alpha u^{2} -v^{2} \tau \frac{du}{dz} +v\tau h\mu \right]=\frac{d^{2} }{dz^{2} } \left(\mu -\eta v\frac{du}{dz} \right)-h\mu  ,  \end{equation}
\begin{equation} \label{2.6} \mu =-\frac{d^{2} u}{dz^{2} } +\left(u-a_{1} \right)\left(u-a_{2} \right)\left(u-a_{3} \right) .  \end{equation}

At $\pm \infty $ all derivative terms equal to zero, so only solutions connecting the stationary states and turning to zero the polynomial in the right-hand side of \eqref{2.6} are possible. If we look for the solution connecting the state $a_{1} $ at $+\infty $ with the state $a_{3} $ at $-\infty $, the suitable form of the Ansatz is:
\begin{equation} \label{2.7} \frac{du}{dz} =\kappa \left(u-a_{1} \right)\left(u-a_{3} \right)=\kappa \left(u^{2} -up+q\right) .  \end{equation}
Here we introduced notations $p=\left(a_{1} +a_{3} \right);\, \, q=a_{1} a_{3} $ for brevity; $\kappa $ is presently an unknown constant, $\kappa >0$ for $\frac{du}{dz} <0$ (`anti-kink'), and $\kappa <0$ for $\frac{du}{dz} >0$ (`kink'). Using \eqref{2.7} the expression for the chemical potential \eqref{2.6} could be transformed,
\begin{equation} \label{2.8} \mu =\left[\frac{1}{\kappa } \left(1-2\kappa ^{2} \right)u+\frac{1}{\kappa } \left(\kappa ^{2} p-a_{2} \right)\right]\frac{du}{dz} . \end{equation}
Alternatively,
\begin{equation} \label{2.9} \mu =-\frac{d}{dz} \left[\frac{du}{dz} -\frac{1}{2\kappa } \left(u^{2} -2a_{2} u\right)\right]. \end{equation}
Substituting \eqref{2.9} into \eqref{2.5} and integrating once we get
\begin{eqnarray} \label{2.10} && {\frac{d}{dz} \left[\mu -\left(\eta v+v\tau h\right)\frac{du}{dz} +\frac{v\tau h}{\kappa } \left(\frac{1}{2} u^{2} -a_{2} u\right)\right]+\left(h-v^{2} \tau \right)\frac{du}{dz}  } \nonumber \\ &&{+\left(\alpha -\frac{h}{2\kappa } \right)u^{2} +\left(\frac{ha_{2} }{\kappa } +v\right)u+C=0}.\end{eqnarray}

Using expression \eqref{2.8} for $\mu $ we transform \eqref{2.10} into
\begin{eqnarray} \label{2.11} && {\frac{1}{\kappa } \frac{d}{dz} \left\{\left[\left(1-2\kappa ^{2} \right)u+\left(\kappa ^{2} p-a_{2} -\eta \kappa v-v\kappa \tau h\right)\right]\frac{du}{dz} +\frac{v\tau h}{2} u^{2} -v\tau ha_{2} u\right\}}\nonumber  \\ &&{+\left(h-v^{2} \tau \right)\frac{du}{dz} +\left(\alpha -\frac{h}{2\kappa } \right)u^{2} +\left(\frac{ha_{2} }{\kappa } +v\right)u+C=0}  .\end{eqnarray}
Inspection of the powers of $u$ in \eqref{2.11} makes evident that for this equation to be satisfied identically for arbitrary$u$, the expression under the derivative should be linear in $u$. This means that
\begin{equation} \label{2.12} 2\kappa ^{2} -1=0, \end{equation}
\begin{equation} \label{2.13} \kappa \left(p-2a_{2} \right)=\eta v .\end{equation}
Then \eqref{2.11} becomes
\begin{equation} \label{2.14)} \left[\left(h\eta -1\right)v^{2} \tau +\frac{\alpha }{\kappa } \right]\kappa u^{2} +\left[\kappa p\tau v^{2} \left(1-h\eta \right)+\left(1-h\eta \right)v\right]u=0 .\end{equation}
Here we used $C$ to cancel $u$-independent terms. Equating to zero coefficients we get two additional constraints
\begin{equation} \label{2.15} \frac{\alpha }{\kappa } =v^{2} \tau \left(1-h\eta \right) ,\end{equation}
\begin{equation} \label{2.16} \left(1-\eta h\right)\left(\kappa pv\tau +1\right)v=0;\, \,  .\end{equation}

If the constraints \eqref{2.12}--\eqref{2.13}, and \eqref{2.15}--\eqref{2.16} are fulfilled, the solution of \eqref{2.7} is simultaneously solution of \eqref{2.5}--\eqref{2.6}. Integrating \eqref{2.7} and taking the maximal steepness point for $z=0$, we get
\begin{equation} \label{2.17)} u=\frac{a_{1} +a_{3} }{2} -\frac{a_{3} -a_{1} }{2} \tanh\left(\frac{1}{2} \kappa \left(a_{3} -a_{1} \right)z\right) .\end{equation}

\setcounter{equation}{0}
\section{The parametric dependence of the solution}\label{s3}

There are four constraints \eqref{2.12}-\eqref{2.13}, \eqref{2.15}-\eqref{2.16} but only two unknowns $\kappa ,\, v$. That is, a constant-velocity traveling wave could be generated by the joint action of two mechanisms only if the two additional constraints are imposed on the system parameters. From \eqref{2.13}, it follows
\begin{equation} \label{3.1} v=\frac{\kappa }{\eta } \left(p-2a_{2} \right). \end{equation}
However, in writing the latter equation we implicitly presumed $\eta \ne 0$, and before proceeding we need to check, what happens if $\eta =0$. We get from \eqref{2.13}, \eqref{2.15} and \eqref{2.16}:
\begin{equation} \label{3.2)} a_{1} +a_{3} =2a_{2} ;\, \, \, \, \frac{\alpha }{\kappa } =v^{2} \tau ;\, \, \, \, \left(\kappa pv\tau +1\right)v=0. \end{equation}
So the solution is possible only if the additional constraint is imposed on the stationary states. If $\alpha =0,\, \, \tau \ne 0$, then $v=0$, i.e., only the static solution is possible. If $\tau =0$ it should be necessary $\alpha =0$, and it follows again $v=0$. If $\alpha \ne 0,\, \, \tau \ne 0$, then $v=-\frac{1}{\kappa p\tau } $. The traveling wave is possible if in addition to $a_{1} +a_{3} =2a_{2} $, there is the constraint $\alpha \tau p^{2} =2\kappa $ ($\left|\kappa \right|=\frac{1}{\sqrt{2} } $, but we keep $\kappa $ to point out that its sign should coincide with the sign of $\alpha $).

Now, if the viscosity $\eta $ is non-zero, eliminating $v$ from \eqref{2.15}-\eqref{2.16}, we get two constraints imposed on the parameters of the system
\begin{equation} \label{3.3} 2\alpha \eta ^{2} =\kappa \tau \left(p-2a_{2} \right)^{2} \left(1-h\eta \right), \end{equation}
\begin{equation} \label{3.4} \left(1-\eta h\right)\left[p\tau \left(p-2a_{2} \right)+2\eta \right]\left(p-2a_{2} \right)=0 .\end{equation}

The latter equation is satisfied if:

1) $p-2a_{2} =0$; then from \eqref{3.3} it should be $\alpha =0$, and from \eqref{3.1} $v=0$; i.e. only the static solution is possible;

2) $1-\eta h=0$; again from \eqref{3.3} it should be $\alpha =0$; however there is no limitation on $p-2a_{2} $, so the traveling wave is possible, and the velocity $v$ is given by \eqref{3.1}. In this case the memory time $\tau $ drops out from the constraints, so there is no influence of memory. Naturally, this case corresponds to the problem without memory, which was considered in \cite{38} (the notations in the present paper are different);

3) $p\tau \left(p-2a_{2} \right)=-2\eta $; or
\begin{equation} \label{3.5} \, \left(p-2a_{2} \right)=-\frac{2\eta }{p\tau } . \end{equation}
Substitution of \eqref{3.5} into the second constraint \eqref{3.3} yields
\begin{equation} \label{3.6} \, \alpha p^{2} \tau =2\kappa \left(1-h\eta \right). \end{equation}

If $\eta \ne 0$, the expression \eqref{3.1} for the velocity is valid. As discussed above, if $\eta =0$ the equations \eqref{3.5} and \eqref{3.6} become
\begin{equation} \label{3.7)} a_{1} +a_{3} =2a_{2} ;\, \, \, \, \, \, \alpha p^{2} \tau =2\kappa . \end{equation}
And the velocity is
\begin{equation} \label{3.8)} v=-\frac{1}{\kappa p\tau } . \end{equation}

In the initial notations the dimensional velocity $\bar{v}$ and the constraints are

For the case 2):
\begin{equation} \label{3.9} \bar{v}=\frac{\kappa \sqrt{\varepsilon \rho } }{\bar{\eta }} \left(\bar{a}_{1} +\bar{a}_{3} -2\bar{a}_{2} \right) ,\end{equation}
\begin{equation} \label{3.10} \bar{\eta }M_{2} =1;\, \, \, \alpha =0 .\end{equation}

For the case 3):

if $\eta \ne 0$ the velocity is still given by\eqref{3.9}; if $\eta =0$
\begin{equation} \label{3.11} \bar{v}=-\frac{\sqrt{\varepsilon } }{\kappa \sqrt{\rho } \left(\bar{a}_{1} +\bar{a}_{3} \right)\tau '} . \end{equation}

The constraints are generally
\begin{equation} \label{3.12} \left(\bar{a}_{1} +\bar{a}_{3} \right)\left(\bar{a}_{1} +\bar{a}_{3} -2\bar{a}_{2} \right)\rho =-\frac{2\bar{\eta }}{\tau '} , \end{equation}
\begin{equation} \label{3.13}
\, \left(\bar{a}_{1} +\bar{a}_{3} \right)^{2} \sqrt{\frac{\rho }{\varepsilon } } =\frac{2\kappa }{\bar{\alpha }\tau '} \, \left(1-\bar{\eta }M_{2} \right). \end{equation}

Summing up, for the conserved and non-conserved dynamics to be `cooperative', the two additional constrains \eqref{3.3}-\eqref{3.4} should be satisfied. There are three essentially different cases when these constraints are satisfied. If a symmetry condition $a_{1} +a_{3} =2a_{2} $ is imposed on the stationary states of the potential, and the convective term is absent, only the static solution is possible.

On the other hand, if there is a link \eqref{3.10} between two kinetic parameters, viscosity $\bar{\eta }$ and kinetic coefficient $M_{2} $, and again $\alpha =0$, the traveling wave exists with velocity \eqref{3.9}. Using \eqref{3.10}, \eqref{3.9} could be rewritten as
\begin{equation} \label{3.14} \bar{v}=\kappa \sqrt{\bar{\varepsilon }\rho } M_{2} \left(\bar{a}_{1} +\bar{a}_{3} -2\bar{a}_{2} \right). \end{equation}

Remarkably, while the dependence of the velocity on the values of stationary states is exactly the same, as for the well known solution of the diffusion equation with cubic nonlinearity, i.e., on the Allen-Cahn part, the coefficient in \eqref{3.14} depends on $\bar{\varepsilon }$ and $\rho $, i.e., on the Cahn-Hilliard part of \eqref{1.1}. It evidently reminds the appearance of the surface tension in the expression for the velocity for the `motion by mean curvature', though the `driving forces' are essentially different. It is well known that for the classic Cahn-Hilliard equation only exact static solutions are possible. Nevertheless, the introduction of the convective and viscous terms, i.e., the appearance of the balance between the external field and dissipation, enables existence of exact one- and two-wave solutions \cite{13}. However, for the convective-viscous CH-AC equation the situation is essentially different. In the absence of memory the very presence of external field makes the existence of an exact constant-velocity traveling wave solution impossible. On the other hand, the existence of such a solution is ensured by the balance between the dissipation and kinetics, see \eqref{3.10}.

In the third case, the constraints are \eqref{3.12}-\eqref{3.13}. Evidently, the very existence of this case is due to the memory. The both constraints exhibit the balance between the `Cahn-Hilliard-` and `Allen-Cahn-parameters' with essential presence of the characteristic memory time $\tau '$. However, for the non-zero viscosity $\eta $ the expression for the velocity is the same. Only if $\eta =0$, the velocity becomes essentially different, see \eqref{3.11}. In this case the dissipation is absent, and the balance between conserved and non-conserved dynamics is due to the memory.


{\small \topsep 0.6ex

}


\begin{thebibliography}{99}

\bibitem{1} Katsoulakis, M.A. \& Vlachos, D.G., Mesoscopic modeling of surface processes. In: Abdallah, N.B., et al. (eds.), Dispersive transport equations and multiscale models, The IMA Volumes in Mathematics and its Applications, 2004, \textbf{136}, Springer, New York, NY, \doi{10.1007/978-1-4419-8935-2_12}.

\bibitem{2} Thiele, U., Thin film evolution equations from (evaporating) dewetting liquid layers to epitaxial growth. J. Phys.: Condens. Matter, 2010, \textbf{22}, no. 8, 084019, \doi{10.1088/0953-8984/22/8/084019}.

\bibitem{3} Cahn J.W. \& Hilliard J.E., Free energy of nonuniform systems. I. Interfacial free energy, J. Chem. Phys., 1958, \textbf{28}, 258--267, \doi{10.1063/1.1744102}.

\bibitem{4} Cahn, J.W., On spinodal decomposition, Acta Metall., 1961, \textbf{9}, 795--801, \doi{10.1016/0001-6160(61)90182-1}.

\bibitem{5} Novick-Cohen, A., The Cahn-Hilliard equation. In: Handbook of Differential Equations, Evolutionary Equations. V.4, Dafermos C.M., Feireisl E. (eds.),  Elsevier B.V., 2008, 201--228, \doi{10.1016/S1874-5717(08)00004-2}.

\bibitem{6} Miranville, A., The Cahn-Hilliard equation and some of its variants, AIMS Mathematics, 2017, \textbf{2}, No.\~3, 479--544, \doi{10.3934/Math.2017.2.479}.

\bibitem{7} Novick-Cohen, A. \& Segel, L.A.,  Nonlinear aspects of the Cahn--Hilliard equation, Physica D, 1984, \textbf{10}, 277--298, \doi{10.1016/0167-2789(84)90180-5}.

\bibitem{8} Novick-Cohen, A., On the viscous Cahn-Hilliard equation, In: Material Instabilities in Continuum Mechanics and Related Mathematical Problems, (J. M. Ball, Ed.), Oxford Univ. Press, Oxford, 1988, 329--342.

\bibitem{9} Leung, K., Theory on morphological instability in driven systems,  J. Stat. Phys., 1990, \textbf{61}, 345-364, \doi{10.1007/BF01013969}.

\bibitem{10} Witelski, T.P., Shocks in nonlinear diffusion, Appl. Math. Lett., 1995, \textbf{8}, 27--32, \doi{10.1016/0893-9659(95)00062-U}.

\bibitem{11} Emmott, C.L. \&  Bray, A.J., Coarsening dynamics of a one-dimensional driven Cahn-Hilliard system, Phys. Rev. E, 1996, \textbf{54}, No.\~5, 4568--4575, \doi{10.1103/PhysRevE.54.4568}.

\bibitem{12} Witelski, T.P., The structure of internal layers for unstable nonlinear diffusion equations, Stud. Appl. Math., 1996, \textbf{96}, 277--300, \doi{10.1002/sapm1996973277}.

\bibitem{13} Mchedlov-Petrosyan, P.O., The convective viscous Cahn-Hilliard equation: Exact solutions, European Journal of Applied Mathematics, 2016, \textbf{27}, 42--65, \doi{10.1017/S0956792515000285}.

\bibitem{14} Mchedlov-Petrosyan,~P.O. \& Davydov,~L.N., Cahn-Hilliard model with Schl\"ogl reactions: interplay of equilibrium and non-equilibrium phase transitions. I. Travelling wave solutions, Condensed Matter Physics, 2020, \textbf{23}, No.~3, 33602: 1--17, \doi{10.5488/CMP.23.33602}..

\bibitem{15} Mchedlov-Petrosyan,~P.O. \& Davydov,~L.N., Cahn-Hilliard model with Schl\"ogl reactions: interplay of equilibrium and non-equilibrium phase transitions. II. Memory effects, Condensed Matter Physics, 2025, \textbf{28}, No.~1, 13601: 1--12, \doi{10.5488/cmp.28.13601}..

\bibitem{16} Allen,~S.M., \& Cahn,~J.W., A microscopic theory for antiphase boundary motion and its application to antiphase domain coarsening, Acta Metallurgica, 1979, \textbf{27}, 1085--1095, \doi{10.1016/0001-6160(79)90196-2}.

\bibitem{17} Imbihl, R., Ertl, G., Oscillatory kinetics in heterogeneous catalysis, Chem.Rev., 1995, \textbf{95}, 697--733, \doi{10.1021/cr00035a012}.

\bibitem{18} Karali, G. \& Katsoulakis, M.A., The role of multiple microscopic mechanisms in cluster interface evolution, J. Differential Equations, 2007, \textbf{235}, 418--438, \doi{10.1016/j.jde.2006.12.021}.

\bibitem{19} Karali, G. \& Ricciardi, T., Existence and asymptotics for a Cahn-Hilliard/Allen-Cahn parabolic equation KURENAI, Kyoto University Research Information Repository, 2009, 1628: 87-100.

\bibitem{20} Karali, G. \& Ricciardi, T., On the convergence of a fourth order evolution equation to the Allen-Cahn equation, Nonlinear Anal., 2010, \textbf{72}, 4271-4281, \doi{10.1016/j.na.2010.02.003}.

\bibitem{21} Israel, H., Miranville, A., \& Petcu, M., Well-posedness and long time behavior of a perturbed Cahn-Hilliard system with regular potentials, Asymptot. Anal., 2013, \textbf{84}, 147--179, \doi{10.3233/ASY-131172}.

\bibitem{22} Karali, G. \& Nagase, Y., On the existence of solution for a Cahn-Hilliard/Allen-Cahn equation, Discrete Contin. Dyn. Systems Ser. S, 2014, \textbf{7}, 127-137, \doi{10.3934/dcdss.2014.7.127}.

\bibitem{23} Antonopoulou, D.C., Karali, G., \& Tzirakis, K., Layer dynamics for the one dimensional $\varepsilon$-dependent Cahn-Hilliard/Allen-Cahn equation, Calc. Var., 2021, \textbf{60}, 207, \doi{10.1007/s00526-021-02085-4}.

\bibitem{24} Hildebrand, M. \& Mikhailov, A.S., Mesoscopic modeling in the kinetic theory of adsorbates, J. Phys. Chem., 1996, 100, 19089--91, \doi{10.1021/jp961668w}.

\bibitem{25} Katsoulakis, M.A. \& Vlachos, D.G., From microscopic interactions to macroscopic laws of cluster evolution, Phys. Rev. Letters, 2000, \textbf{84}, 1511--14, \doi{10.1103/PhysRevLett.84.1511}.

\bibitem{26} Vlachos, D.G. \& Katsoulakis, M.A., Derivation and validation of mesoscopic theories for diffusion-reaction of interacting molecules, Phys. Rev. Lett., 2000, \textbf{85}, 3898--3901, \doi{10.1103/PhysRevLett.85.3898}.

\bibitem{27} Horntrop D.J., Katsoulakis M.A., \& Vlachos D.G., Spectral Methods for Mesoscopic Models of Pattern Formation, Journal of Computational Physics, 2001, \textbf{173}, 364--390, \doi{10.1006/jcph.2001.6883}.

\bibitem{28} Mitlin, V.S., Dewetting of solid surface: Analogy with spinodal decomposition. J. Colloid Interface Sci., 1993, \textbf{156}, 491--497, \doi{10.1006/jcis.1993.1142}.

\bibitem{29} Ji, H. \& Witelski, T.P., Steady states and dynamics of a thin-film-type equation with non-conserved mass, Euro. Journal of Applied Mathematics, 2020, \textbf{31}, 968--1001, \doi{10.1017/S0956792519000330}.

\bibitem{30} Fort, J. \& Mendez, V., Wavefronts in time-delayed reaction-diffusion systems. Theory and comparison to experiment, Rep. Progr. Phys., 2002, \textbf{65}, 895--954, \doi{10.1088/0034-4885/65/6/201}.
\bibitem{31} Galenko, P., Phase-field model with relaxation of the diffusion flux in nonequilibrium solidification of a binary system, Phys. Lett. A, 2001, \textbf{287}, 190--197, \doi{10.1016/S0375-9601(01)00489-3}.

\bibitem{32} Debussche, A., A singular perturbation of the Cahn-Hilliard equation. Asymptotic Anal., 1991, \textbf{4}, no. 2, 161--185. \doi{10.3233/ASY-1991-4202}.

\bibitem{33} Galenko, P. \& Jou, D., Diffuse-interface model for rapid phase transformations in nonequilibrium systems, Phys. Rev. E, 2005, \textbf{71}, 046125, \doi{10.1103/PhysRevE.71.046125}.

\bibitem{34} Galenko, P. \& Lebedev, V., Non-equilibrium effects in spinodal decomposition of a binary system. Phys. Lett. A, 2008, \textbf{372}, 985--989, \doi{10.1016/j.physleta.2007.08.070}.

\bibitem{35} Gatti, S., Grasselli, M., Miranville, A., \& Pata, V., On the hyperbolic relaxation of the one-dimensional Cahn-Hilliard equation, J. Math. Anal. Appl., 2005, \textbf{312}, 230--247, \doi{10.1016/j.jmaa.2005.03.029}.

\bibitem{36} Gatti, S., Grasselli, M., Pata, V., \& Miranville, A., Hyperbolic relaxation of the viscous Cahn-Hilliard equation in 3-D, Math. Models Methods Appl. Sci., 2005, \textbf{15}, 165-198, \doi{10.1142/S0218202505000327}.

\bibitem{37} Folino, R., Lattanzio, C., \& Mascia, C., Slow dynamics for the hyperbolic Cahn-Hilliard equation in one space dimension, Math. Meth. Appl. Sci., 2019, \textbf{42}, 2492--2512, \doi{10.1002/mma.5525}.

\bibitem{38} Mchedlov-Petrosyan, P.O. \& Davydov, L.N., Convective viscous Cahn-Hilliard/Allen-Cahn equation: exact solutions, 2020, arXiv:2002.08747[physics.comp-ph], \doi{10.48550/arXiv.2002.08747}.

\end{thebibliography}
\end{document}